\documentclass[11pt]{article}

\usepackage{epsfig, amsfonts, amsmath, amssymb, amsthm, pictex, verbatim, amsthm, pdfsync, mathrsfs}
\usepackage{tensind}
\usepackage[round]{natbib}
\usepackage{endnotes}
\usepackage[normalem]{ulem}

\usepackage{hyperref}  %This package should often be loaded last, to prevent conflict with other packages.
\hypersetup{colorlinks=true, citecolor=blue, urlcolor=blue,linkcolor=blue} 
\usepackage{graphicx}
\usepackage{amsmath}
\usepackage{cases}
\usepackage{amssymb}
\usepackage{hyphenat}
\usepackage[hypcap]{caption} %When hyperrefs link to figures, they go to the top of the figure, instead of to the caption.

\usepackage{tikz} 					%Drawing pictures within LaTeX.
\usetikzlibrary[patterns]				% Different patterns to fill areas.
\usetikzlibrary[arrows]				% Different arrow heads
\usetikzlibrary{positioning}			% Rotating (e.g. vertical axis labels)
\usepackage[colorinlistoftodos, textwidth=2.8cm, shadow]{todonotes}				%Inserting todo notes.  %Change colour of bublles and remove lines: \todo[noline,color=red]{comment}
\usepackage{marginnote}

\usepackage{framed}			%The package creates three environments:framed, which puts an ordinary frame box around the region,shaded, which shades the region, and left­bar, which places a line at the left side. The environments allow a break at their start (the \FrameCommand enables creation of a title that is “attached” to the environment); breaks are also allowed in the course of the framed/shaded matter.

\usepackage{comment}

\usepackage{booktabs} % for much better looking tables
\usepackage{multirow}
\usepackage{tabularx}
\usepackage{fixltx2e} % Allows for \textsubscript
\usepackage{booktabs} % for much better looking tables
\usepackage{multirow}

\usepackage{verbatim}  % Required for the automatic word count.

\usepackage{caption}
\usepackage{subcaption}   % These two packes (caption and subcaption) are used to get nice arrays of figures.

\usepackage{cite}  % [1,2,3]  will become [1-3].

\usepackage[margin=1.52in]{geometry}
\usepackage{authblk}

\usepackage{xparse}

\definecolor{cblue}{RGB}{100,5,255}    
\definecolor{cred}{RGB}{155,50,40} 
\definecolor{cgreen}{RGB}{5,165,20}  
\definecolor{corange}{rgb}{1.0,0.49,0.0}  
                                                                   
\title{Integrating Dark Matter, Modified Gravity, and the Humanities}
\author[1,2,3]{Niels C.M.\ Martens\footnote{\url{nmartens@uni-bonn.de}}}
\author[1,4]{Miguel \'{A}ngel Carretero Sahuquillo}
\author[1,4,5]{Erhard Scholz}
\author[1,2]{Dennis Lehmkuhl}
\author[1,3]{Michael Kr\"{a}mer}
\affil[1]{DFG Research Unit ``The Epistemology of the Large
	Hadron Collider'' (grant FOR 2063)}
\affil[2]{Lichtenberg Group for History and Philosophy of Physics, University of Bonn, Germany}
\affil[3]{Institute for Theoretical Particle Physics and Cosmology, RWTH Aachen University, Germany}
\affil[4]{Interdisciplinary Centre for Science and Technology Studies, University of Wuppertal, Germany}
\affil[5]{Subdivision for Teachers Education and History of Mathematics, Faculty of Mathematics and Natural Sciences, University of Wuppertal, Germany}
\date{\small Editorial of a special issue on dark matter \& modified gravity, distributed across \emph{Studies in History and Philosophy of Modern Physics} and \emph{Studies in History and Philosophy of Science}. Published version of the open access editorial (in SHPS) available here: \href{https://doi.org/10.1016/j.shpsa.2021.08.015}{https://doi.org/10.1016/j.shpsa.2021.08.015}. The six papers are collected here: \href{https://www.sciencedirect.com/journal/studies-in-history-and-philosophy-of-science-part-b-studies-in-history-and-philosophy-of-modern-physics/special-issue/10CR71RJLWM}{https://www.sciencedirect.com/journal/studies-in-history-and-philosophy-of-science-part-} \href{https://www.sciencedirect.com/journal/studies-in-history-and-philosophy-of-science-part-b-studies-in-history-and-philosophy-of-modern-physics/special-issue/10CR71RJLWM}{b-studies-in-history-and-philosophy-of-modern-physics/special-issue/10CR71RJLWM}.}
\begin{document}
	
\maketitle

\begin{quote}
	There are different sorts of conflicts between theories. One familiar kind of conflict is that in which two or more theorists offer rival solutions of the same problem. In the simplest cases, their solutions are rivals in the sense that if one of them is true, the others are false. More often, naturally, the issue is a fairly confused one, in which each of the solutions proffered is in part right, in part wrong and in part just incomplete or nebulous. There is nothing to regret in the existence of disagreements of this sort. Even if, in the end, all the rival theories but one are totally demolished, still their contest has helped to test and develop the power of the arguments in favour of the survivor. 
	
	\hfill --Gilbert Ryle (\citeyear[p.1]{ryle1954}) 
\end{quote}

\section{Introduction} \label{intro}

%\niels{Michael \& co, could you please help me by providing relevant citations for this first section? Should we just cite some review papers? Or all the classical papers? I'm worried that there won't be an end to it once we start citing all the relevant papers.}

%\todo{maybe cite: \url{http://philsci-archive.pitt.edu/18810/}}

%\todo{add new Merritt paper in Synthese: Feyerabends rule and dark matter}

It is well known that the conjunction of the following two assumptions fails spectacularly at accounting for cosmological and astrophysical data:
\begin{enumerate}
	\item Most of the matter in the universe is luminous matter, namely the particles (excluding neutrinos) described by the standard model of particle physics (SM);
	\item Their gravitational interaction is correctly described (before reaching the regime of quantum gravity) by General Relativity (GR)---and Newtonian Gravity as its non-relativistic limit.
\end{enumerate}  
Discrepancies show up at a large range of scales: in galaxies, galaxy clusters and at the cosmological scale. This poses a dilemma. Option 1 gives up on the first assumption by postulating an additional kind of matter, which---since we do not `see' it---must be non-luminous, i.e.,\ dark matter, and potentially even extends the SM with additional interactions (between dark matter and luminous matter and/or between dark matter particles themselves). Option 2 modifies General Relativity, and, potentially, even its non-relativistic limit.\footnote{A third, non-revisionary option would be to have standard model neutrinos make up all of the required dark matter, but their mass and cosmic density are known to be insufficient for these neutrinos to solve the problem all by themselves. Beyond-the-standard-model neutrinos are still considered to be a possible solution; they fall under option 1.}

The majority of contemporary physicists has taken the first horn of the dilemma, as encapsulated in the standard model of cosmology, $\Lambda$CDM, which has as its ingredients General Relativity, dark energy ($\Lambda$) and cold dark matter (i.e.\ dark matter moving at non-relativistic speeds). This model has earned its spurs at large scales: at the cosmological scale, but also at the level of galaxy clusters. Adding an amount of dark matter to the universe with a total mass several times that of the luminous matter allows us to model structure formation from density fluctuations in the early universe as progressing to the extent that we observe it today. The resulting characteristic imprint on the angular power spectrum of the cosmic microwave background (CMB) is consistent with this. Moreover, additional dark matter in galaxy clusters can correct for the fact that galaxies move faster within clusters than expected based on only the gravitational pull of the luminous matter in the cluster. 

Galaxy rotation curves can be fitted by specifying a dark matter halo profile for each individual galaxy. However, this does not naturally account for various strong astrophysical correlations observed across galaxies, associated with a single acceleration scale $a_0 = 1.2 \times 10^{-10}$ m s$^{-2}$. Moreover, structure formation simulations typically do not produce the right halo profiles nor the correct amount of halos of each size/type---these are the so-called ``small scale problems''. Finally, if dark matter is a particle, it has so far managed to avoid being detected non-gravitationally, i.e.\ directly (via the recoil interaction with nuclei on earth), indirectly (via annihilation into SM cosmic rays) or by being produced at particle colliders such as the Large Hadron Collider.

A minority of contemporary physicists favours the second horn of the dilemma, and attempts instead to modify the laws of gravity.\footnote{Modifications of standard gravity are considered for a number of reasons; this special issue focuses only on modifications of gravity aimed at accounting for the specific empirical discrepancies mentioned in the main text.} Modified Newtonian Dynamics (MOND) enhances the strength of the gravitational interaction---compared to Newton's gravitational force---at accelerations much smaller than $a_0$. MOND uniquely and correctly predicts the observed galactic correlations as well as the galaxy rotation curves. A range of relativistic extensions has been constructed, but these typically a) fail to account completely for the dynamical discrepancies in galaxy clusters, b) fail to avoid the constraints imposed by the recent observation of electromagnetic and gravitational waves arriving almost simultaneously from binary neutron star merger event GW170817, and/or c) are typically wrong, or more often silent, about cosmological data such as the CMB.

In summary, at larger scales dark matter (DM) seems to currently be more successful than modified gravity (MG), and vice versa at smaller scales. 

\section{Integrated Research Programme}

There are two main motivations for this special issue on interdisciplinary perspectives on (the interface between) dark matter and modified gravity. Firstly, improving the communication between these two camps of physicists in a debate that is somewhat infamous for its polemical nature and lack of effective communication, for a variety of physics-related, funding-related, ideological, sociological and historical reasons. We take the benefits of inter-programme communication, as illustrated in the opening quote, to be fairly straightforward, but they are particularly vividly illustrated within Peter Galison's (\citeyear[Ch.9]{galison1997}) picture of the progress of science. 
%The logical positivists claimed that there is progress in science in virtue of the unification or stability provided by an unbroken, cumulative language of theory-neutral observations. The antipositivists, such as Kuhn, claimed that there are no theory-neutral observations; the language of science is theory. But the radical discontinuity of theory across scientific revolutions paints a disunified, destabilised picture of science in terms of unconnected blocks of theory, threatening the possibility of progress in science. 
A naive picture of science might have it that there is progress in science in virtue of the unification or stability provided by an unbroken, cumulative language of theory-neutral observations. The antipositivists, such as Kuhn, challenged views of this sort by arguing that there are no theory-neutral observations; the language of science is theory. Kuhn emphasised discontinuity of this language of theory across scientific revolutions. But a radical discontinuity of theory across scientific revolutions threatens to paint a disunified, destabilised picture of science in terms of unconnected blocks of theory, jeopardising the possibility of progress in science. 
Galison claims that, once we focus on scientific communities instead of the global language of science, we realise that it is in fact disunification (of those communities) which ensures strength, overall stability and progress. On Galison's intercalated picture of the development of science, different communities, such as instrumental/theoretical/experimental cosmologists/particle-physicists/relativists, form a tight web despite each having their own traditions, cultures, problems and favoured solutions, values, tools, etc. No local revolution/discontinuity within a single community can threaten the stability of the overall web, \emph{as long as it is tight enough}. What is this glue, this mortar, that makes this a strong web rather than isolated individual communities with incommensurable paradigms? Trading zones---zones that allow local coordination between communities without homogenization of values and traditions. The concept of trading zones has not only a metaphorical aspect---eg., the blogosphere---but also a concrete, physical aspect---eg., the way these communities are located within a physics department. Local contact languages---eg., the concept of Feynman diagrams, which is a simple tool that allows communication between communities even if it may have a different value or even meaning for each community involved---ensure that such communication/coordination is possible, without the need for a global language. Contact languages may be a mere jargon, a more complex pidgin, or even a (self-standing) full-blown creole, and could be spoken either by all members of the communities involved or by middle(wo)men---either representatives of each community or third parties---with interactional expertise \citep{collinsevansgorman}. In sum, trade---communication and coordination---between, e.g., DM and MG communities, is crucial for the progress, robustness and strength of science.

We, the interdisciplinary members of the six-year-project ``LHC, dark matter \& gravity'' within the research unit ``Epistemology of the Large Hadron Collider'',\footnote{\url{www.lhc-epistemologie.uni-wuppertal.de}} aim to help improve the barely existing DM/MG trading zone---with respect to both its literal and figurative aspect---in order to move away from an infertile stand-off between incommensurable paradigms. Concretely, by having gathered seventy dark matter physicists, modified gravity physicists, scholars of science and technology studies (STS) as well as historians, philosophers and sociologists of science during the ``Dark Matter \& Modified Gravity Conference" at RWTH Aachen University, Germany, from 6 to 8 February 2019. More abstractly, by editing this virtual special issue in an interdisciplinary journal. 

A second motivation for this special issue arises from the fact that, although the empirical discrepancies underlying the DM/MG debate are considered to be one of the biggest problems in modern physics---in 2019 alone more than 2,300 papers were posted on arxiv.org with ``dark matter'' in the title or abstract---hardly any attention has been paid to it by the humanities. Although physicists have of course told the story of their research programme from an internal perspective (e.g.\ \citet{sandersbook,bertonehooper2018,peebles2020}), there are only three contributions from historians of physics \citep{vanderburgh2014c,deswart2017,deswartforth}, on the early history of the dark matter problem. This leaves ample room for historical scholarship on different aspects of DM, and especially on MG and the DM/MG interface. Sociologists and STS scholars---despite ample work on, e.g., trading zones and interactional expertise \citep{gorman2010}---have not discovered the debate at all.\footnote{But for sociological reflections by physicists, see Stacy McGaugh's blog, \url{www.tritonstation.com}, and \citeauthor{lopez2014} (\citeyear{lopez2014}). One may also keep an eye out for the upcoming PhD thesis in sociology by Adrien de Sutter.} Philosophers of physics have made some more headway---see below for citations\footnote{See also the President's Plenary Symposium on dark matter at the PSA2018 conference: \url{https://spark.adobe.com/page/AiXfAUmLTaEbB/}.}---%\citep{vanderburgh2000,vanderburgh2003,vanderburgh2005,vanderburgh2014,vanderburgh2014b,kosso2013,sus2014,massimipeacock2014,lahavmassimi2014,jacquart2018,jacquartcompanion,merritt2017,merritt2020,boyd2018,gueguenforth,debaerdemaekerforth,kashyapms},\footnote{\cniels{Add other philosophers that I cite below. What to do with physicists who mentioned a little bit of philosophy in their papers?}} 
but this is still only a piece-meal literature.\footnote{This is of course not to say that there are no further connections with the broader literature, e.g.\ the growing literatures on the philosophy of cosmology, on the philosophy of astronomy and on models and simulations, but a disproportionately small amount of attention is typically being paid to dark matter.} Especially given the current standoff (see \hyperref[intro]{\S \ref{intro}}), we contend that this debate within physics can benefit from contributions from the humanities---and vice versa. Adding philosophy, history, sociology and STS communities to Galison's honeycomb structure would only strengthen it. The best proof of this pudding is of course in the eating. This special issue serves as a first taste of this integrative pudding (see \hyperref[SI]{\S\ref{SI}}). 

To get you started on making your own pudding, we offer here not a recipe, but a list of possible ingredients, grouped together in three categories: semantic, descriptive and normative research questions. The list is of particular interest for readers interested in the interface between DM and MG, but also for readers interested in only one of these two approaches. This list is definitely not exhaustive, nor do all questions fall into only one of the three categories---we have listed them under what we take to be their primary category. The aim of this structured list is merely to give an impression of what a rich, integrated research programme on the interface between dark matter and modified gravity might look like, whilst simultaneously allowing us to reference the few scattered publications on this topic almost exhaustively---the other side of this coin being identification of the gaps in the literature. 

\vspace{3mm}
\noindent
{\bf Semantic (and metaphysical) questions}
\begin{itemize}
\setlength\itemsep{0em}
	\item Is dark matter `proper' matter, or does the concept of matter come in degrees? Does the answer depend on whether one considers, e.g., dark matter that interacts non-gravitationally with luminous matter, self-interacting dark matter, or purely gravitating dark matter (cf. sterile neutrinos)? What makes a field a dark matter field? What makes a field a modification of the gravitational field? Is there a strict DM/MG dichotomy, or are there fields that fit into both categories, or perhaps sit more comfortably somewhere in between? What use, if any, does a conceptual DM/MG distinction have? These questions receive particular interest in light of a recent trend towards `hybrid theories' which attempt to be the best of both worlds.\footnote{See either contribution to this special issue by Martens \& Lehmkuhl for a tentative list of hybrid theories.} Could these hybrid theories form a semantic bridge between the DM and MG communities---do they reveal, rather than ``mere mortar'', a full-blown creole/interlanguage that might eventually even lead to an independent community/ research field, similarly to how biochemistry arose from biology and chemistry?\footnote{\citet{collinsevansgorman}; \citet[p.33-34,43]{galison2010}.} 
	\item Do different communities (e.g.\ theoretical gravitational physicists, cosmologists, astronomers, experimental particle physicists) mean the same thing when they talk about dark matter, both now and in the past? If not, is there a common core that is shared between the different uses,\footnote{\citet{debaerdemaekerforth,martensms}.} which may serve as a contact language in the trading zone?
	\item Is MOND a law, a set of axioms, a constraint on the final theory, an algorithm, mere phenomenology,\footnote{\citet{helbig2020}.} or something else? Which of these would facilitate the use of MOND in the trading zone?
	\item Is it an objective fact or convention whether there is dark matter?\footnote{\citet{merritt2017}.} Whether a field is (dark) matter or a modification of gravity?\footnote{\citet{duerr2021}.} And whether it is dark or luminous matter? %(These questions are of particular interest if one thinks that there is a three-way equivalence between f(R) gravity, the Jordan frame representation of Jordan-Brans-Dicke Theory and the Einstein Frame representation of Jordan-Brans-Dicke Theory.)
	\item What would it mean to be a realist about DM---if this is at all possible yet---given the current uncertainty about its nature and the lack of direct, non-gravitational observation?\footnote{\citet{ruphy2011,jacquartcompanion,merritt2021b,martensms}; \citeauthor{allzen2021} (\citeyear{allzen2021}, \citeyear{allzenmsa}, \citeyear{allzenmsc}).}
\end{itemize}

%In one representation of JBD theory, the JBD scalar seems to be dark matter (it only couples to luminous matter (including light) indirectly via the metric) and in the other representation it is luinous matter (i.e. couples universally to luinous matter). Conventionality between dark and luminous matter?

\noindent
{\bf Descriptive questions}
\begin{itemize}
\setlength\itemsep{0em}
	\item Which of the following factors have shaped and are shaping the debate, and how: empirical data, theoretical virtues (e.g.\ explanatory power, non-ad-hocness), robustness, idealisation, philosophical motivations and guiding principles (e.g. various equivalence principles), historical factors, sociological factors?\footnote{\citet{ruphy2011,kosso2013,sus2014,vanderburgh2014,massimipeacock2014,deswart2017,boyd2018,kadowaki2018,cirkovic2018,deswartforth,debaerdemaekerforth}.} How does this debate compare to other historical debates?\footnote{\citet{lahavmassimi2014}.}
	\item How do constraints from cosmology, astronomy and particle physics interact?
	\item What types of underdetermination and inferential circularities, if any, are there in the context of DM and MG?\footnote{\citet{vanderburgh2000,vanderburgh2003,vanderburgh2005,ruphy2011,kosso2013,smeenk2013,smeenkfortha,sus2014,vanderburgh2014,massimipeacock2014,lahavmassimi2014,smeenksep,jacquartcompanion,kashyapms}.}
	\item Is the debate best understood in terms of theories,  models,\footnote{\citet{jacquartphd,massimi2018}.} a Popperian analysis,\footnote{\label{merhelb}\citet{merritt2017,merritt2021}, and a response by \citet{helbig2020}.} (Kuhnian) paradigms,\footnote{\citet[Chps.~8,12]{sandersbook}; \citet{sandersmsold,lahavmassimi2014,mcgaugh2015,merritt2017,jacquartcompanion}.} (Lakatosian) research programmes,\footnote{\citet{merritt2017,merritt2020,lahavmassimi2014}.} (Laudan's) research traditions, a Feyerabendian analysis,\footnote{\citet{merrittfeyerabend}} and/or intercalated communities trading with each other? If DM and MG are Kuhnian paradigms, are they in a crisis or in a period of normal science? If they are Lakatosian research programmes, are they degenerative or progressive? Can tools from the digital humanities confirm that there is a divide between the DM and MG communities and lack of a trading zone; if so, in what sense and why?
\end{itemize}

\noindent
{\bf Normative, epistemological and methodological questions}
\begin{itemize}
\setlength\itemsep{0em}
	\item Where should resources---time, funding,\footnote{\citet[p.20-21]{collinsevansgorman}.} intellect---go? Should we be research programme pluralists, or monists---focusing only on one preferred research programme, or even just a few specific models---and if so, which one(s)?\footnote{\citet{cirkovic2018,bertonetait2018,merritt2020}.} How promising are hybrid theories? What would it take, if anything at all, to abandon DM or MG? If there is underdetermination between MG and DM, how should we break it---which criteria for theory choice are relevant and how do they apply?\footnote{\citet{vanderburgh2000,vanderburgh2014b,ruphy2011,smeenk2013,smeenkfortha,lahavmassimi2014,massimipeacock2014,smeenksep,boyd2018,debaerdemaekerforth}.} %\cniels{Somewhere we need to mention a question that allows referring to Milgrom's (and McGaugh's) contributions to this special issue, such as ``How does the struggle between DM and MG compare to other struggles in the history of science?''}
	\item How does the explanatory power of DM and of MG compare?\footnote{\citet{vanderburgh2014b,jacquartphd}.} What is the epistemic status, robustness and explanatory power of (AI-involving) DM simulations?\footnote{\citet{ruphy2011,massimi2018,gueguenforth}.} Do simplified DM models explain? Are DM and/or MG ad hoc?\footnote{\citet{smeenk2013,massimipeacock2014,merritt2017,merritt2020,helbig2020}.} Are the DM and MG research programmes, theories and models predictive?\footnote{\citet{sandersmsnew,massimi2018,merritt2020}.} Falsifiable?\footnote{See fn.\ref{merhelb}.} What guiding principles, if any, should govern the development of new models? 
	\item Is there a difference, in degree or even in kind, between purely-DM astronomical objects vs.\ luminous astronomical objects vs.\ collider-produced DM particles vs.\ cosmic DM particles reaching the surface of the earth, in terms of a) the directness of their observation (or, more generally, the epistemic access we have to these objects), and b) the logic of justification for/confirmation of their existence?\footnote{\citet{jacquart2018,debaerdemaekerforth}; \citeauthor{allzen2021} (\citeyear{allzen2021}, \citeyear{allzenmsa}).}
	\item Should dark matter be explored independently from dark energy?
	\item What would it take to confirm that a new non-luminous particle discovered at the LHC is \emph{the} cosmic dark matter particle---especially in light of the indirect nature (i.e.\ via missing energy) of the inference of such a particle as well as the inability of the LHC to determine cosmological density and stability?
	\item What are good strategies for integrating dark matter, modified gravity and the humanities---for creating trading zones?\footnote{\citet{gorman2010}.} Could hybrid theories play a role? Should all members of all communities involved speak the contact language, or are middle(wo)men more successful?\footnote{\citet{collinsevansgorman}.} Should those middle(wo)men be physicists, or is this were the humanities come in, or both?
\end{itemize}

\section{Contributions to this Special Issue} \label{SI}

This special issue consists of six papers, four of which appear in \emph{Studies in History and Philosophy of Modern Physics}, and two of which appear in \emph{Studies in History and Philosophy of Science} (SHPS), as the former journal was reabsorbed into the latter in January 2021. 

The first contribution, by \textbf{Mordehai Milgrom}, is titled ``\href{https://www.sciencedirect.com/science/article/abs/pii/S1355219819301972}{MOND vs.\ dark}\linebreak \href{https://www.sciencedirect.com/science/article/abs/pii/S1355219819301972}{matter in light of historical parallels}''. It portrays the debate between dark matter and MOND as one of rivalling paradigms. Milgrom's analysis---falling within the category of descriptive questions mentioned in the previous section---frames the debate by comparing it to and contrasting it with similar historical examples of rivalling paradigms, such as the Copernican and quantum revolutions, the demise of the aether, and the rise of atomism and of general relativity. His analysis includes discussions of prediction vs.\ post hoc explanation, ad-hocness, the emergence of hybrid theories, and MOND-DM discriminants.

\textbf{Siska de Baerdemaeker} and \textbf{Nora Mills Boyd} focus on the normative, methodological question of how to proceed in light of the small scale problems, in their paper ``\href{https://www.sciencedirect.com/science/article/abs/pii/S1355219820301088}{Jump ship, shift gears, or just keep on chugging: Assessing the} \linebreak \href{https://www.sciencedirect.com/science/article/abs/pii/S1355219820301088}{responses to tensions between theory and evidence in contemporary cosmology}''. \linebreak Their bottom-up approach---focusing on scientific practice, in the context of a concrete case study---stands in contrast with the top-down, general approach to understanding theory change familiar from, e.g., Popper and Lakatos. Are the small scale problems anomalies that are sufficiently severe to rationally justify abandoning the dark matter research programme (and maybe even switching to modified gravity)? If not, do they nevertheless force a modification of the cold dark matter hypothesis? Or is the most rational way of proceeding to patiently continue attempting to solve these challenges within the research programme in its standard form? De Baerdemaeker \& Boyd side with this third option, arguing that it is crucial to prioritise obtaining a more comprehensive determination of non-gravitational contributions (to dark matter simulations) from known, baryonic physics.

%\textbf{Siska de Baerdemaeker} and \textbf{Nora Mills Boyd} focus on the normative, methodological question of how to proceed in light of the small scale problems, in their paper \emph{Jump ship, shift gears, or just keep on chugging: Assessing the responses to tensions between theory and evidence in contemporary cosmology}. In contrast to the top-down, general approach to understanding theory change familiar from, e.g., Popper and Lakatos, this paper proceeds in a bottom-up fashion, by considering the specific case study of dark matter's small scale challenges. Are these anomalies that are sufficiently severe to rationally justify abandoning the dark matter research programme (and maybe even switching to modified gravity)? If not, do they nevertheless force a modification of the cold dark matter hypothesis? Or is the most rational way of proceeding to patiently continue attempting to solve these challenges within the research programme in its standard form? De Baerdemaeker \& Boyd side with this third option, arguing that it is crucial to prioritise obtaining a more comprehensive determination of non-gravitational contributions (to dark matter simulations) from known, baryonic physics. 

In the category of semantic questions we find two companion papers by \textbf{Niels Martens} and \textbf{Dennis Lehmkuhl}.\footnote{For reasons of objectivity, editorial management of their contributions was left entirely to the Editors-in-Chief of \emph{Studies in History and Philosophy of Modern Physics}.} What makes a new field either a modification of the gravitational field/ spacetime structure, or a (dark) matter field? Could it be both? Neither? Their first companion paper, ``\href{https://www.sciencedirect.com/science/article/pii/S135521982030109X}{Dark Matter = Modified Gravity?} \linebreak      
\href{https://www.sciencedirect.com/science/article/pii/S135521982030109X}{Scrutinising the spacetime--matter distinction through the modified gravity/ dark} \linebreak
\href{https://www.sciencedirect.com/science/article/pii/S135521982030109X}{matter lens}'', focuses on a specific hybrid theory, Berezhiani \& Khoury's `superfluid dark matter theory', as a case study. The new field in this theory is argued to be both fully a dark matter field and (in the superfluid regime) fully a modification of gravity/spacetime. The second companion paper, ``\href{https://www.sciencedirect.com/science/article/pii/S1355219820301106}{Cartography of the space}\linebreak \href{https://www.sciencedirect.com/science/article/pii/S1355219820301106}{of theories: An interpretational chart for fields that are both (dark) matter and}\linebreak \href{https://www.sciencedirect.com/science/article/pii/S1355219820301106}{spacetime}'', provides a chart consisting in three groups of possible interpretations for such theories, to help navigate the single space of theories---as opposed to one space of dark matter theories and a separate space of modified gravity theories associated with two rivalling paradigms.

This leaves the two contributions that are published in the SHPS journal. \textbf{Stacy McGaugh}'s descriptive paper---``\href{https://www.sciencedirect.com/science/article/pii/S0039368121000728}{Testing galaxy formation and dark matter with}\linebreak \href{https://www.sciencedirect.com/science/article/pii/S0039368121000728}{low surface brightness galaxies}''---complements Milgrom's contribution by zooming in on galaxy formation. McGaugh relays the history of observations of low surface brightness galaxies. Emphasis is placed on the need for auxiliary hypotheses like feedback to save the dark matter hypothesis in light of these observations. Increasing the complexity of the models often resembles squeezing a toothpaste tube: solving one (fine-tuning) problem (re)creates another (fine-tuning) problem (contra De Baerdemaeker \& Boyd). While complex models including dark matter may be able to accommodate the data, this is not the same as predicting the data \emph{a priori} (cf.\ Milgrom). The main question this leaves us with, i.e.\ the main puzzle for $\Lambda$CDM, is: Why does MOND get (m)any prediction(s) right?

\textbf{Melissa Jacquart}'s paper---``\href{https://www.sciencedirect.com/science/article/abs/pii/S003936812100100X}{$\Lambda$CDM and MOND: A debate about models or}\linebreak \href{https://www.sciencedirect.com/science/article/abs/pii/S003936812100100X}{theory?}''---belongs in the category of descriptive questions. Whereas the received view portrays the debate between $\Lambda$CDM and MOND as a disagreement between theories, Jacquart argues---building upon recent work by Massimi (\citeyear{massimi2018})---that the relationship between these two approaches is better understood when analysed in terms of models. Models can be extended not only to different scale level domains, but also in terms of purpose: description, prediction or explanation. $\Lambda$CDM and MOND are each successful in their originally intended domain of application---the large scale and the meso/galactic scale, respectively---with respect to their originally intended purposes, i.e.\ description and prediction in both cases. The problem arises once we realise that justifying the extension of each of these models to a different domain is connected to explanatory fit.

\section*{Acknowledgements}

We would like to acknowledge support from the DFG Research Unit ``The Epistemology of the Large Hadron Collider'' (grant FOR 2063). We are grateful to the speakers at the ``Dark Matter \& Modified Gravity Conference'' in Aachen, to all those who submitted a paper to this special issue, and to the anonymous referees for their excellent work. We would also like to thank Wayne Myrvold for his continued support throughout the process, and Chris Smeenk for help with the bibliography.

\bibliography{bibsidmmgarxiv}

\end{document}